\pdfoutput=1
\documentclass[prb,reprint,showpacs,superscriptaddress]{revtex4-1}
\usepackage{graphicx}
\usepackage{amsfonts}
\usepackage[figuresright]{rotating}
\usepackage{float}
\usepackage{amssymb}
\usepackage{amsmath}
\usepackage{hyperref}

\def\be{\begin{equation}}
\def\ee{\end{equation}}
\def\bea{\begin{eqnarray}}
\def\eea{\end{eqnarray}}

\def\nn{\nonumber}

\begin{document}

\title{Time-reversal Invariant SU(2) Hofstadter Problem in Three Dimensional Lattices}
\author{Yi Li}
\affiliation{Princeton Center for Theoretical Science, Princeton University, Princeton, New Jersey 08544, USA}
\date{June 2, 2015}

\begin{abstract}
We formulate the three-dimensional $SU$($2$) Landau level
problem in cubic lattices with time-reversal invariance.
By taking a Landau-type $SU$($2$) gauge, the system can be reduced into
one-dimension, as characterized by the $SU$($2$) generalization of the usual Harper equations
with a periodic spin-dependent gauge potential.
The surface spectra indicate the spatial separation of helical states
with opposite eigenvalues of a lattice helicity operator.
The band topology is investigated from both the analysis of
the boundary helical Fermi surfaces and the calculation of
the $\mathbb{Z}_2$-index based on the bulk wave functions.
The transition between a three-dimensional weak topological insulator to a strong
one is studied as varying the anisotropy of hopping parameters.
\end{abstract}
\pacs{73.43.Cd, 71.70.Ej, 75.70.Tj}
\maketitle

\section{Introduction}
\label{sec:intro}
The study of topological states of matter has become a major research
focus of contemporary condensed matter physics \cite{qi2010,qi2011,Hasan2010}.
An early example is the two-dimensional (2D) quantum Hall effect based on the Landau
level quantization of electrons in the magnetic
field, whose topological band structure is characterized by the first
Chern number \cite{klitzing1980,thouless1982,halperin1982,
kohmoto1985,haldane1988}.
Its chiral edge modes give rise to dissipationless charge transport
and quantized Hall conductance.
In the past decade, tremendous progress has been made on the
time-reversal invariant topological insulators in both 2D and 3D,
whose band structure topology is described by the $\mathbb{Z}_2$ index
\cite{kane2005,bernevig2006a,bernevig2006,qi2008,fu2007,moore2007,roy2010}.
At the boundary, the 2D and 3D topological insulators exhibit 1D
helical edge modes
and 2D helical surface modes, respectively.
Various systems of the 2D and 3D topological insulators
have been observed experimentally through various
spectroscopic and transport measurements \cite{bernevig2006,konig2007,
hsieh2008,zhang2009,xia2009,chen2009,qu2010}.
Another important
development is based on high-dimensional generalizations of Landau levels.
Landau level wave functions explicitly exhibit elegant analytic
properties, which played an important role in the study
of 2D fractional quantum Hall effects \cite{laughlin1983,haldane1983}.
Zhang and Hu \cite{zhang2001} pioneered the study of Landau levels
on four- and higher-dimensional spheres and other compact manifolds by coupling
fermions to gauge potentials of non-Abelian monopoles
\cite{karabali2002,elvang2003,bernevig2003,hasebe2010,edge2012,
hasebe2014,hasebe2014a}.

Recently, Landau levels (LLs) in three- and higher dimensional flat continuum space have also
been discussed \cite{li2013,li2013a,haaker2014}.
In 3D and 4D, their Hamiltonians describe spin-$\frac{1}{2}$ fermions in
an SU(2) gauge potential, which were constructed from harmonic oscillators
with certain spin-orbit coupling (SOC).
The corresponding Landau level wave functions are the same as those of harmonic
oscillators, but they are reorganized by the SOC term
to exhibit non-trivial topology.
In the symmetric-type $SU$($2$) gauge, the 3D and 4D rotational symmetries
are explicitly maintained with exactly flat energy spectra and degenerate angular momenta.
The lowest Landau level wave functions exhibit interesting quaternion analyticity\cite{li2013}.
In the Landau-type gauge, they behave like spatially separated
2D helical Dirac fermion modes or 3D chiral Weyl fermion modes.
Modes with opposite helicities or chiralities are shifted
along the third or fourth dimension in opposite directions,
respectively \cite{li2013a}.
These LLs have also been generalized to high dimensional Dirac fermions
and parity-breaking systems \cite{li2012,li2012d}.
One issue of the topological states based on high-dimensional
Landau levels is how to define their topological index.
The non-trivial topology of filled Landau levels was illustrated
through the effective boundary Hamiltonians, which exhibit helical Fermi surfaces \cite{li2013,li2013a,li2012}.
However, unlike the 2D 
case, there was no full 
translation symmetry for the 3D Landau level Hamiltonians, and
thus the $\mathbb{Z}_2$-index defined based on the Bloch wave function
structure could not be applied.

A lattice construction of the Hofstadter Hamiltonian for the 3D
LL problem will be helpful to directly calculate
the topological index of that system with translation symmetry imposed.
For the 2D case, the lattice version of the LL problem
based on the $U(1)$ vector field is well known as the Hofstadter problem
\cite{hofstadter1976}.
In the Landau gauge, the $U(1)$ vector potential becomes a periodic scalar potential in the reduced 1D Harper equation.
The non-trivial Chern number is interpreted as the linking number
between fundamental loops on the complexified energy surface
with the topology of a Riemann surface \cite{hatsugai1993a,hatsugai1993}.
Recently, this Hofstadter Hamiltonian was realized experimentally
in optical lattices by laser-assisted
tunneling \cite{miyake2013,aidelsburger2013,celi2014,jaksch2003}.
Also, Hofstadter problems have been theoretically generalized
to systems with non-Abelian gauge groups \cite{osterloh2005,kimura2012,scheurer2014}.

In this article, we will consider the 3D LLs of the Landau-type SU(2)
gauge in which the system shows translation symmetry in the
$xy$ plane.
For each pair of in plane momenta $(k_x,k_y)$,
the system can be reduced to one dimension, as described by
the SU(2) Harper equation exhibiting the
periodic configuration of spin-dependent potential.
The bulk and surface spectra are explicitly calculated, and
the non-trivial band topology are analyzed through the
surface helical Fermi surface and also from the analysis
on the parity eigenvalues of the bulk wave functions.
Topological band structure transitions between the weak
and strong 3D topological insulators are studied.

The rest of this paper is organized as follows:
In Sect. \ref{sect:lattice}, the lattice model for 3D
LLs in the SU(2) Landau gauge and the corresponding
SU(2) Harper equation are constructed.
In Sect. \ref{sect:surface}, the topological band structure
is analyzed through the helical surface spectra
and the $\mathbb{Z}_2$-index based on
the parity eigenvalue analysis on the bulk wave function.
In Sect. \ref{sect:transition}, the topological band structure
transition in studied.
Discussions on the relationship with the previous work on the 3D
Hofstadter problem in the tilted magnetic field and experimental
realizations are presented in Sect. \ref{sect:disc}.
Conclusions are summarized in Sect. \ref{sect:conclusions}.

\section{3D Landau levels on a cubic lattice and the 1D $SU(2)$ Harper equation}
\label{sect:lattice}
In this section, we construct the 3D $SU(2)$ Landau level in lattices with
full three-dimensional translation symmetry, and we show that it
can be reduced into a family of generalized 1D $SU$($2$) Harper equations.

\subsection{The 3D Landau level in the continuum}
\label{sect:continuum}
The 2D Landau level in the magnetic field has been generalized into 3D continuum, which was defined as a spin-$\frac{1}{2}$ electron
coupling to an external SU(2) vector potential $\vec A$ and a scalar
potential $V(x)$ \cite{li2013,li2013a}
as
\begin{eqnarray}
H_{3DLL}=\frac{(\vec{p}-\frac{e}{c}\vec A(\vec r))^2}{2m}+V(\vec r).
\label{eq:ham_cont}
\end{eqnarray}
In the Landau-type gauge, $\vec A(\vec r)$ and $V(\vec r)$ only depend
on the $z$-coordinate, which can be chosen as
\bea
A_x(\vec r)&=&G \sigma_y z, \ \ \, A_y(\vec r)= -G \sigma_x z,
\ \ \, A_z=0; \nn \\
V(\vec r)&=&-\frac{1}{2}m \omega^2 z^2,
\label{eq:landaugauge}
\eea
where $G$ is a coupling constant, $\sigma_{x,y,z}$ are Pauli matrices, and $\omega=eG/(mc)$.
Equation (\ref{eq:ham_cont}) describes a 3D topological insulator
possessing the TR symmetry and translational symmetry along the $x$-$y$ plane.
At an open boundary perpendicular to the $z$ axis, each filled
Landau level contributes a helical Dirac Fermi surface
\cite{li2013}. The surface states carrying opposite helicities are spatially separated at
$z>0$ and $z<0$ surfaces respectively.
This can be shown explicitly by expanding Eq. (\ref{eq:ham_cont}) as
$H_{3DLL}=p_z^2/(2m) + 1/2 m \omega^2
[z- l_{so}^2 k_{2d} \hat{\Sigma}_{2d} (\hat k_{2d})]^2$,
where  $\vec k_{2d}=(k_x,k_y)$ and $\hat \Sigma_{2d}$ is the helicity operator defined as
$\hat \Sigma_{2d} (\hat k_{2d} )=\hat k_x \sigma_y -\hat k_y \sigma_x$ with $\hat k_{x,y}=k_{x,y}/k_{2d}$.
Alternatively, Eq. (\ref{eq:ham_cont}) is also equivalent to an electron
in a quantum-well with $z$-dependent spin-orbit coupling strength as
\bea
H_{3DLL}&=&
\frac{\vec p^2}{2m}+\frac{1}{2}m \omega^2 z^2
-\omega z (p_x \sigma_y-p_y \sigma_x).
\label{eq:ham_cont2}
\eea

\begin{figure}[htbp]
\centering
\includegraphics[width=0.6\linewidth]{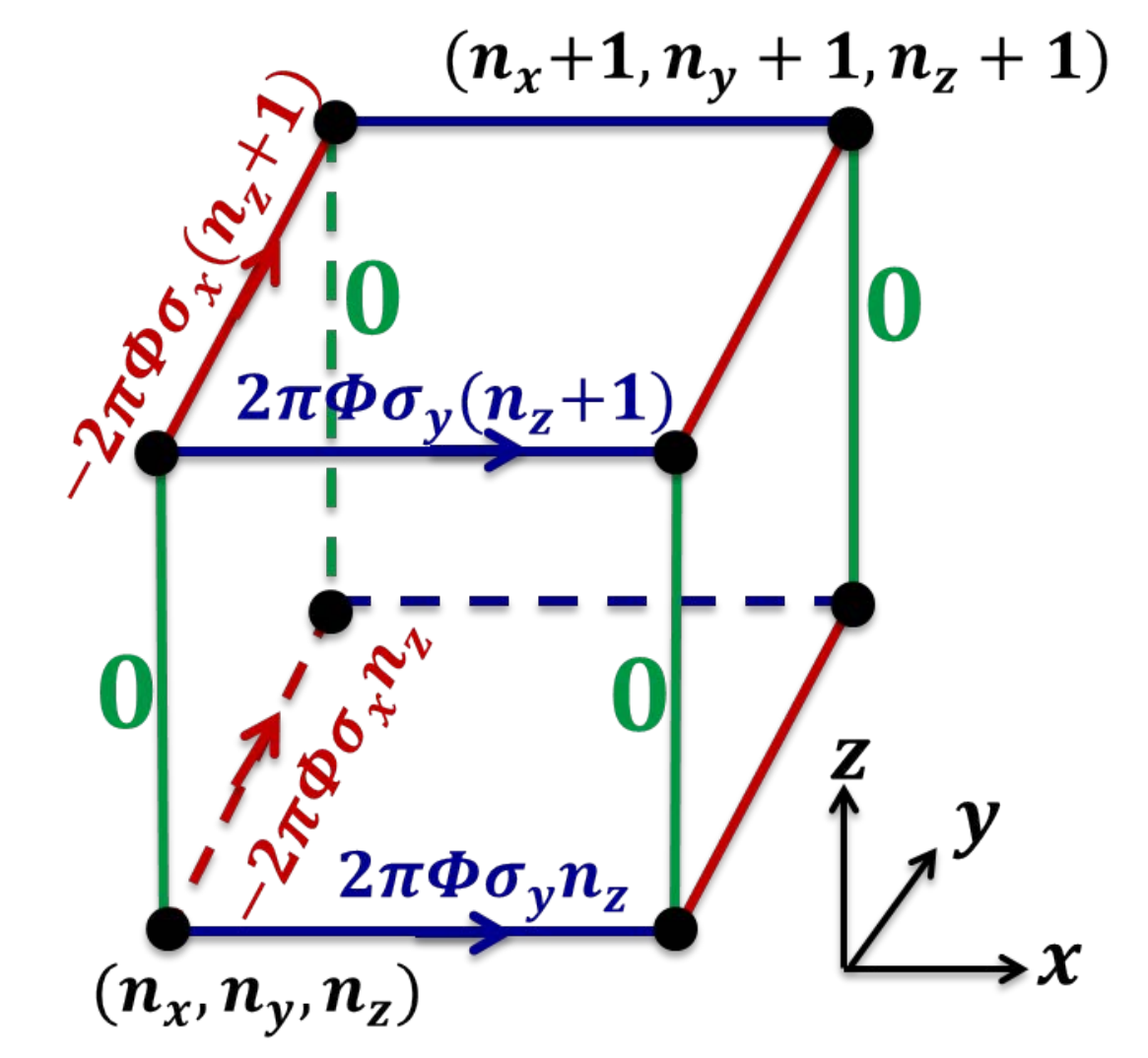}
\caption{(Color online) The configuration of Landau-type $SU(2)$ gauge
fields $\vec A(n_x, n_y, n_z)$ defined on the bonds of the cubic lattice,
which has coordinate dependence only on $n_z$.
}
\label{fig:gauge}
\end{figure}

\subsection{The SU(2) Hofstadter problem in the 3D cubic lattice}
Now let us construct the lattice version of the 3D LL Hamiltonian
Eq. (\ref{eq:ham_cont}), or, Eq. (\ref{eq:ham_cont2}) as a tight-binding
model on a cubic lattice.
The kinetic energy becomes the nearest-neighbor hopping term;
the SO coupling is realized as the $SU(2)$ gauge potential
defined on each bond of the cubic lattice, and the scalar quadratic potential is converted as a periodic potential along the
$z$ axis on the lattice.
The lattice Hamiltonian reads
\bea
H_{latt}&=& -\sum_{a,\vec n,s,s^\prime} t_a
\Big \{c^{\dagger}_{\vec n + \hat e_a , s} [e^{i 
A_{\vec n + \hat e_a, \vec n}} ]_{ss'} c_{\vec n, s'} \nn \\
&+& h.c. \Big\} +
\sum_{\vec n, s} V(n) c^{\dagger}_{\vec n, s} c_{\vec n, s},
\label{eq:3dlattice}
\eea
where $c_{\vec n, s}$ ($c^{\dagger}_{\vec n, s}$) is the annihilation
(creation) operator of electrons at site $\vec n = (n_x, n_y, n_z)$
with spin $s=\uparrow,\downarrow$.
$\hat e_a$ and $t_{a}$ ($a=x,y,z$) are, respectively, the lattice unit vector and
the hopping amplitude along the $a$-direction.
The previous $SU(2)$ Landau-type gauge of Eq. (\ref{eq:landaugauge}) defined in the continuum
can now be defined on each bond of the cubic lattice as follows:
\bea
A_{\vec n + \hat x, \vec n}&=& \frac{2\pi\Phi}{\Phi_0} \sigma_y n_z, \ \ \,
A_{\vec n + \hat y, \vec n}= -\frac{2\pi\Phi}{\Phi_0} \sigma_x n_z,\nn \\
A_{\vec n + \hat z, \vec n}&=&0,
\eea
where $\Phi_0=e/(hc)$ is the flux quantum, whose value is taken as 1 in this paper.
As illustrated in Fig. \ref{fig:gauge}, this gauge configuration can
be obtained by rotating a 2D quantum spin-Hall (QSH) \cite{wu2006} Hofstadter problem
defined in the $xz$ plane by 90$^{\circ}$ around its $z$ axis, so that its $yz$ plane
forms another QSH problem. The QSH problem at each plane can be understood as two Kramerss copies of the 2D U(1) Hofstadter problem with opposite fluxes for opposite spins. Further, because of the SOC, the spin quantization axis along $y$ in the $xz$ plane QSH problem is also rotated by 90$^{\circ}$, to be along $x$, in the $yz$ plane QSH problem, which imposes the 3D feature of the topological insulator and reflects the four-fold SO coupled rotation symmetry in the lattice.
For later convenience, we choose the $n_z$-dependent scalar potential as
\bea
V(n_z)=2t_y \cos (2\pi \frac{\Phi}{\Phi_0} n_z).
\eea
$A_{\vec n + \hat e_a, \vec n}$ and $V(n_z)$ are independent of $n_x$
and $n_y$, hence, Eq. (\ref{eq:3dlattice})
explicitly maintains translational symmetry along the $xy$ plane.
In addition, it also possesses time-reversal and parity symmetries.


Next we transform the lattice Hamiltonian Eq. (\ref{eq:3dlattice}) into
the 1D Harper equation.
Because of the translation symmetry in the $xy$ plane, the in plane
momenta $k_x$ and $k_y$ are good quantum numbers.
For a given set of values of $k_x$ and $k_y$, we introduce
the partial Fourier transform as
\bea
c^\dagger_{n_z,s}(k_x,k_y)=\frac{1}{\sqrt{L_xL_y}}
\sum_{n_x,n_y} e^{-i k_x n_x-i k_y n_y} c^\dagger_{\vec n, s}.
\eea
Then Eq. \ref{eq:3dlattice} is reduced to
\bea
H_z(k_x,k_y)&=&-t_z \sum_{n_z;s}\Big\{c^{\dagger}_{n_z+1;s}(k_x,k_y)
c_{n_z;s}(k_x,k_y) +h.c. \Big\} \nn \\
&-&2\sum_{n_z,s,s^\prime} c^{\dagger}_{n_z;s}(k_x,k_y)
V_{n_z;s,s'} c_{n_z;s'}(k_x,k_y) \nn \\
\label{eq:harper}
\eea
in which
\bea
V_{n_z;s,s'} &=& \delta_{ss'} r(k_x,k_y) \cos (2 \pi n_z \frac{\Phi}{\Phi_0}) \nn \\
&+& [-i\tilde{k}_- \sigma_+ +i \tilde{k}_+ \sigma_-]_{ss'}
\sin (2 \pi n_z \frac{\Phi}{\Phi_0}),
\label{eq:spin_incom}
\eea
and
\bea
r(k_x,k_y)&=&t_x \cos k_x + t_y(\cos k_y -1), \nn \\
\tilde{k}_{\pm}&=& t_x \sin k_x \pm i t_y \sin k_y,  \nn \\
\sigma_{\pm}&=&\frac{1}{2}(\sigma_x \pm i   \sigma_y).
\eea

Eq. (\ref{eq:harper}) consists of the usual
hopping along the $z$-direction and the periodic onsite
spin-dependent potential Eq. (\ref{eq:spin_incom}).
In comparison, the Harper equation for the 2D Landau level problem
in the Landau gauge is a hopping Hamiltonian in the presence
of a periodic onsite scalar potential.
As a result, Eq. (\ref{eq:harper}) maintains time-reversal symmetry
as
\bea
H_z(k_x,k_y)=T H_z(-k_x,-k_y) T^{-1}
\eea
with $T=i\sigma_2 K$ and $K$ is the complex conjugation.

\begin{figure}[htbp]
\centering
\centering\includegraphics[width=0.6\linewidth]{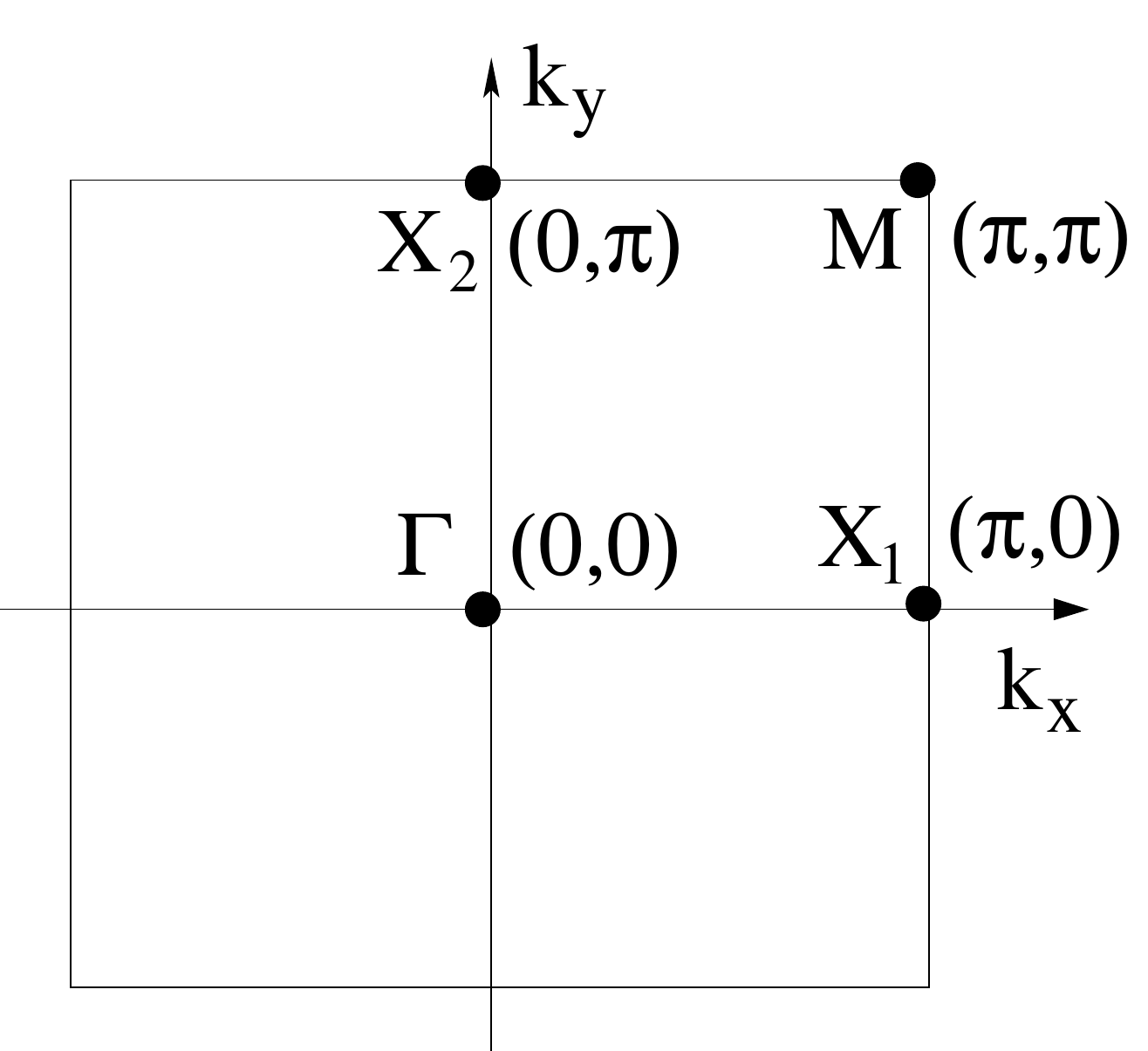}
\caption{The 2D surface Brillouin zone and the time-reversal
invaraint points $\Gamma=(0,0)$, $X_1=(\pi,0)$, $X_2=(0,\pi)$,
and $M=(\pi,\pi)$.  }
\label{fig:BZ}
\end{figure}

\subsection{The SU(2) transfer matrix}
We consider the case of $\Phi/\Phi_0=p/q$ with $p$ and $q$ coprime integers,
and then $H_z(k_x,k_y)$ becomes
periodical along the $z$-direction with an enlarged unit cell size of $q$.
Below we also assume that the lattice size along the $z$-direction $L_z$
is an integer multiple of $q$ as $L_z=lq$.
On a single particle basis with momenta $k_x$ and $k_y$,
\bea
| \Psi (k_x,k_y)\rangle = \sum_{n_z;s} \Psi_{n_z;s}(k_x,k_y)
c^{\dagger}_{n_z;s}(k_x,k_y) |0\rangle,
\eea
the corresponding Harper equation becomes
\bea
&-&[\Psi_{n_z+1;s}(k_x,k_y) + \Psi_{n_z-1;s}(k_x,k_y)]\nn \\
&-&\frac{2}{t_z} \sum_{s'} V_{n_z;s,s'} \Psi_{n_z;s'}(k_x,k_y)
=\epsilon \Psi_{n_z;s}(k_x,k_y),
\label{eq:harper_2}
\eea
where $\epsilon = E/t_z$.
In the form of transfer matrix, Eq. (\ref{eq:harper_2}) is
represented as
\bea
\left[
    \begin{array}{c}
    \Psi_{n_z+1,s}(\epsilon, k_x,k_y) \\
    \Psi_{n_z,s}(\epsilon, k_x,k_y) \\
    \end{array}
\right]
&=& T_{n_z;ss^\prime}(\epsilon, k_x,k_y)
\nn \\
&\times&
\left[
    \begin{array}{c}
    \Psi_{n_z,s'}(\epsilon, k_x,k_y) \\
    \Psi_{n_z-1,s'}(\epsilon, k_x,k_y) \\
    \end{array}
\right].
\eea
The transfer matrix is defined as
\bea
T_{n_z;ss^\prime}(\epsilon, k_x,k_y)&=&
\left(
  \begin{array}{cc}
    {\cal E}_{ss^\prime} & -I_{ss^\prime} \\
    I_{ss^\prime} & 0 \\
  \end{array}
\right),
\label{eq:transfer_1}
\eea
in which $I$ is the $2\times 2$ identity matrix;
and
\bea
{\cal E} &=&-\epsilon I
-\frac{2}{t_z} r(k_x,k_y)  \cos (2 \pi \Phi n_z)\nn \\
&-& \frac{2}{t_z} \sin (2\pi\Phi n_z)
(-i \tilde{k}_- \sigma_+ + i \tilde{k}_+ \sigma_-).
\eea

Noticing that the spin-dependence of the transfer matrix $T_{n_z;ss^\prime}$
is $n_z$-independent, thus the Harper equation (Eq. (\ref{eq:harper_2})) can be
decoupled into two sets in the ``lattice helicty'' eigen-basis.
The lattice helicty operator is defined as
\bea
\Sigma_{\textrm{L}}= t_x \sin k_x \sigma_y - t_y \sin k_y \sigma_x,
\eea
which can be viewed as the lattice generalization of the
helicity operator $\Sigma=\hat k_x \sigma_y - \hat k_y \sigma_x$
in the continuum.
Its eigenvalues are $\pm \sqrt{t_x^2\sin^2 k_x +t_y^2\sin^2 k_y}$,
corresponding to the in plane eigenstates
\bea
\chi_{\pm} (k_x, k_y)
=\frac{1}{\sqrt 2}\left(
   \begin{array}{c}
     \mp i \frac{t_x \sin k_x -i t_y\sin k_y}{\sqrt{t_x^2\sin^2 k_x +t_y^2\sin^2 k_y}} \\
     1 \\
   \end{array}
 \right).
\eea
Under the time-reversal transformation, $\Sigma_L$ as well as its
eigenstates are invariant.
In this representation, the transfer matrix $T_{n_z;ss^\prime}$
is decomposed into two sets of $T_{n_z;\pm}$ with the $2\times 2$
form for $\Sigma_L$ positive and negative helicities, respectively,
as
\bea
T_{n_z;\pm}(\epsilon, k_x,k_y)&=&
\left(
  \begin{array}{cc}
    {\cal E_\pm} & -1 \\
    1 & 0 \\
  \end{array}
\right),
\label{eq:transfer_2}
\eea
with
\bea
{\cal E_\pm} &=&-\epsilon
-\frac{2}{t_z} r(k_x,k_y) \cos (2 \pi \Phi n_z)\nn \\
&\mp& \frac{2}{t_z} \sin (2\pi\Phi n_z) \sqrt{t^2_x\sin k_x^2+
t^2_y\sin k_y^2}.
\eea

\begin{figure}[htbp]
\centering
\centering\includegraphics[width=0.49\linewidth]{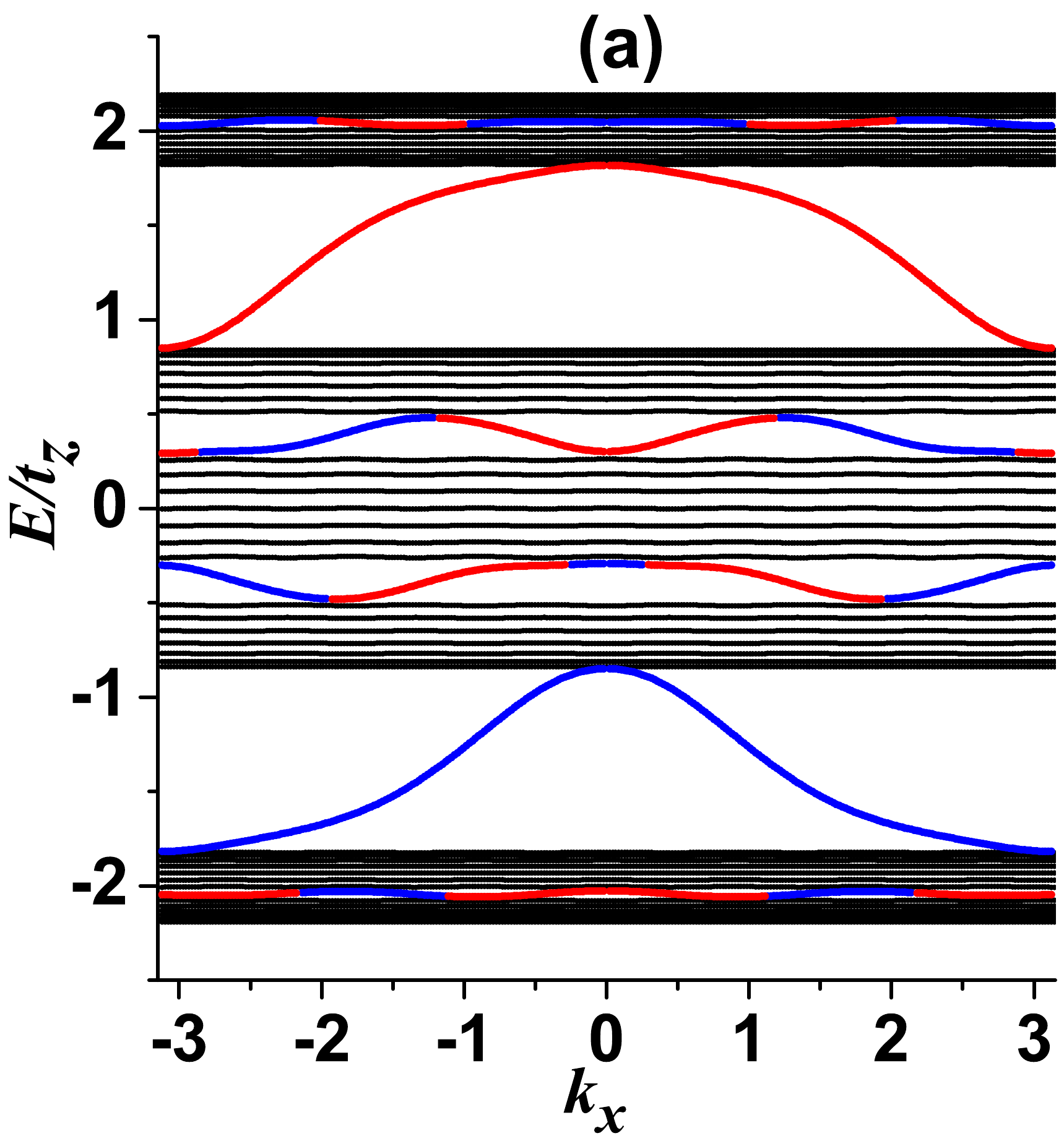}
\centering\includegraphics[width=0.49\linewidth]{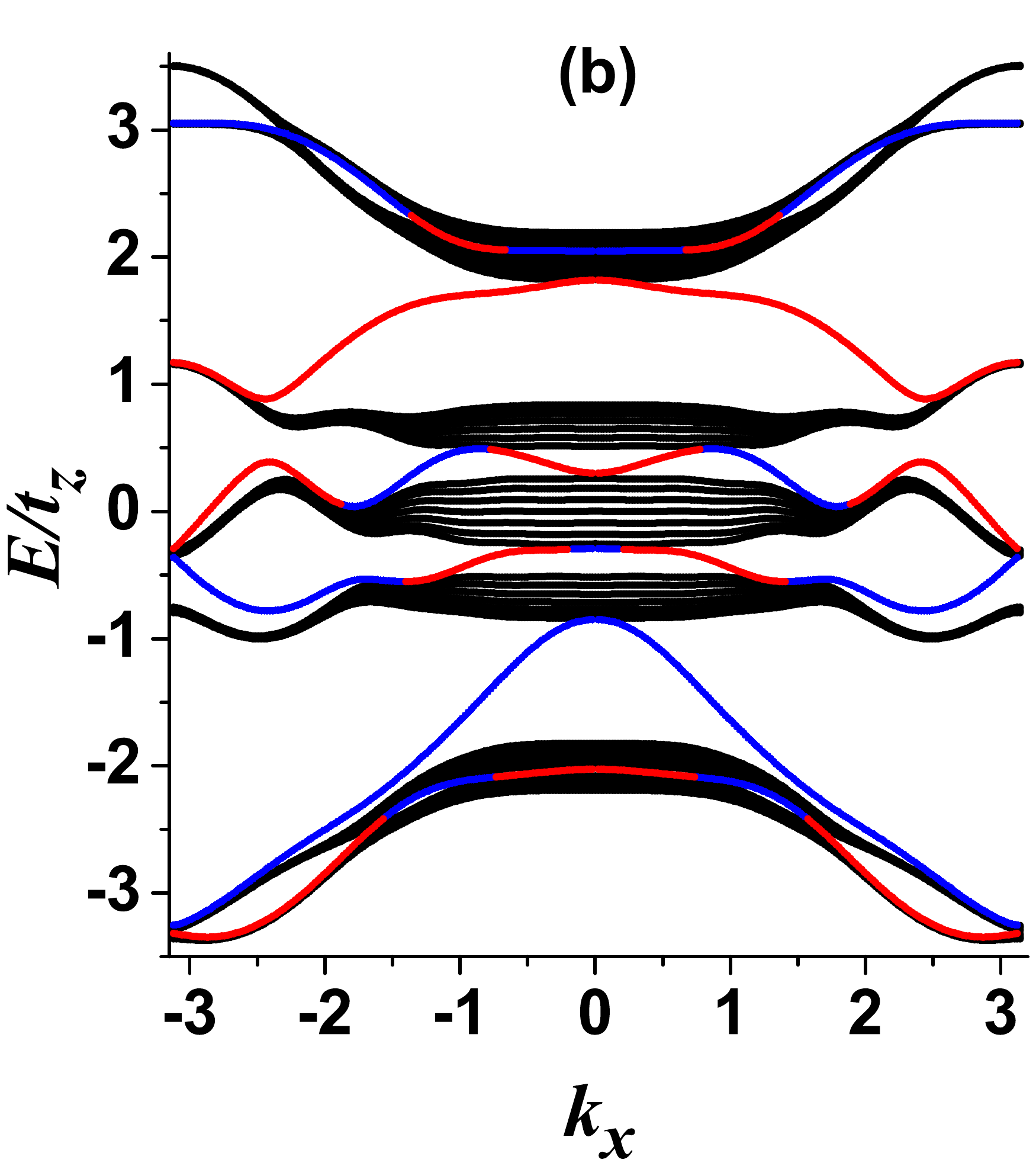}
\centering\includegraphics[width=0.49\linewidth]{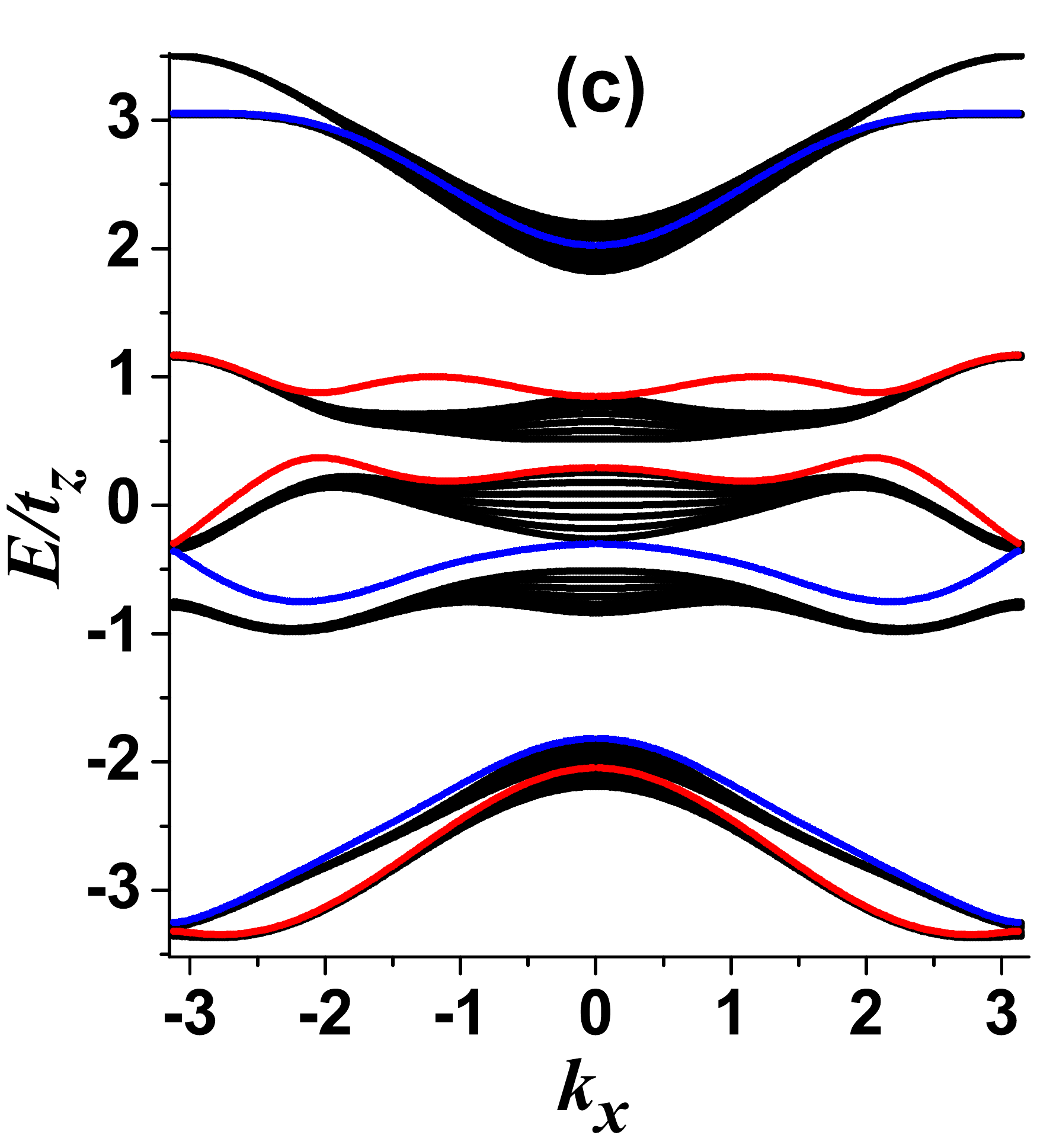}
\caption{The bulk and surface spectra along the high symmetry lines connecting time-reversal invariant points in the Brillouin zone.
($a$) $X_1$-$\Gamma$-$X_1$, ($b$) $M$-$\Gamma$-$M$,
($c$) $M$-$X_2$-$M$, respectively.
The red and blue lines represent the surface states
with positive and negative helicities, respectively.
Here, only surface modes localized at the lower boundary are plotted, and those of the upper surface
are of the same values in the energy spectrum but with opposite lattice helicities.
The following parameters are used: $t_x=t_y=0.5$, $t_z=1$,  $p/q=2/7$,
and $L_z=56$.
}
\label{fig:spectra}
\end{figure}

Following the method introduced in Ref. [\onlinecite{hatsugai1993}],
we define the matrix $M^\pm(\epsilon,k_x,k_y)$ as the product of
$T_{i;\pm}(\epsilon, k_x,k_y)$ for sites $i$'s in a unit cell with $q$-sites as
\bea
M^\pm(\epsilon,k_x,k_y)&=& \prod_{i=1}^q T_{i;\pm}(\epsilon, k_x,k_y).
\label{eq:M_matrix}
\eea
For the open boundary condition, the energy of the surface states
can be determined by the zeros of the matrix elements
$M_{21}^\pm(\epsilon,k_x,k_y)$ which are the roots of ($q-1$)th-order
polynomial equations.
For each lattice helicity sector, the $q-1$ roots $\mu_j^\pm (j=1,2,.., q-1)$
denote the surface states in the $q-1$ gaps between adjacent bands.

\begin{figure}[htbp]
\vspace{5mm}
\centering
\centering\includegraphics[width=0.45\linewidth]{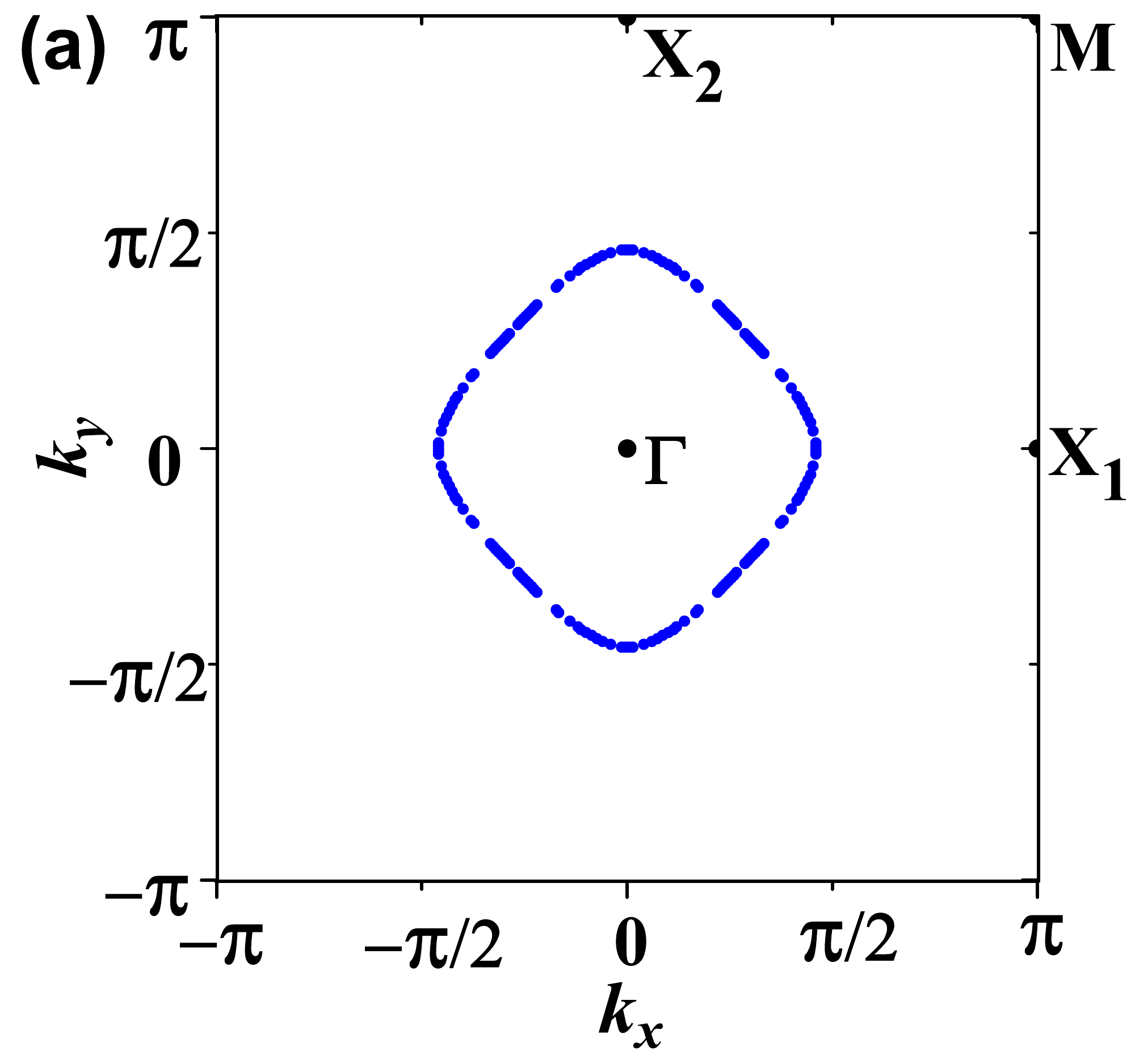}
\centering\includegraphics[width=0.45\linewidth]{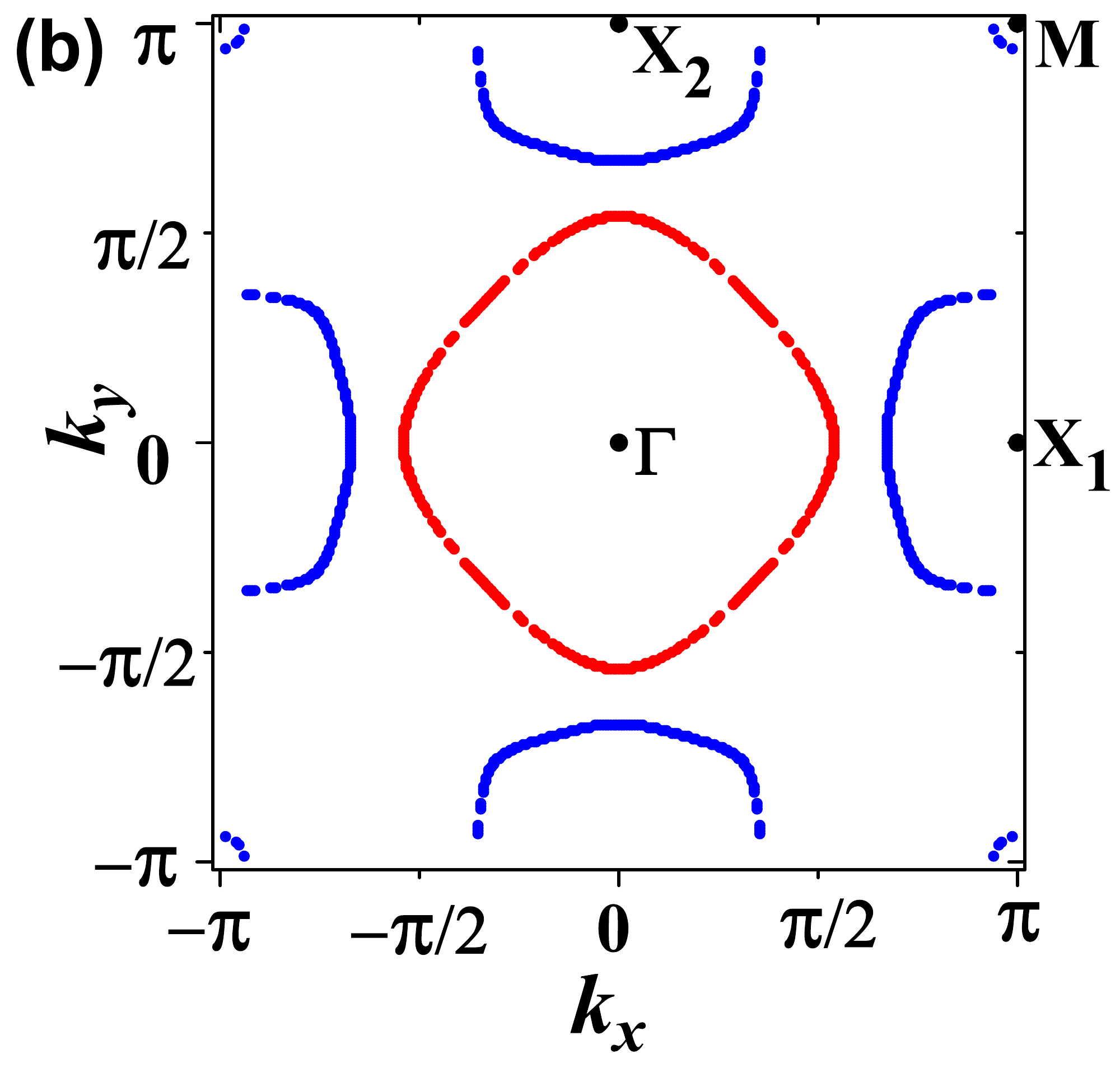}
\caption{(Color online)
The boundary Fermi surface for states localized at the lower surface.
Here, blue and red colors represent the negative and positive helicity
states respectively.
(a) $E_F=-1.5 t_{\parallel}$ lying in the 2nd gap. Only one
boundary helical Fermi surface appears.
(b) $E_F=-0.47 t_{\parallel}$ lying the 3rd gap. There are four
boundary helical Fermi surfaces.
Parameter values are $t_x=t_y=0.5$, $t_z=1$, $p/q=2/7$, and $L_z=56$.
}
\label{fig:surfacecircle}
\end{figure}

\section{The band topology with the tetragonal symmetry}
\label{sect:surface}
In this section, we present the numeric results of the surface and bulk energy spectra of the 3D Hofstadter problem, as well as the analysis of $Z_2$ topological index from the parity eigenvalues of the bulk wave functions.
We first consider the case with tetragonal symmetry in the
$xy$ plane.

\subsection{The energy spectra}
The Harper equation (Eq. (\ref{eq:harper})) is solved numerically with respect to
$k_x$ and $k_y$ in the 2D Brillouin zone as depicted in Fig. \ref{fig:BZ}.
A typical flux value of $\Phi=p/q=2/7$ is used: Each unit cell contains
7 sites, thus the spectra consist of 7 bands.
To fully open band gaps, we choose an anisotropic ratio between
$t_z$ and $t_{x,y}$ as $t_x/t_z=t_y/t_z=0.5$.
The bulk and surface spectra are plotted along different high symmetry lines connecting time-reversal invariant points in the Brillouin zone
in Fig. \ref{fig:spectra} ($a$), ($b$), and ($c$), respectively.
In each gap between adjacent bands, helical surface states appear
with their helicities marked with different colors in
Fig. \ref{fig:spectra}.

The states on the upper and lower surfaces can be distinguished
by the values of $M^\pm_{11}(\epsilon)$ where the $\epsilon$ is
a surface state energy.
If $|M^\pm_{11}(\mu_j)|<1 ~(>1)$, then the corresponding eigenstates
are localized on the lower $(n_z=1)$ or upper $(n_z=L_z)$ boundaries,
respectively \cite{hatsugai1993}.
If $|M^\pm_{11}(\mu_j)|=1$, it means that the corresponding states
merge into bulk states, and they are no longer considered surface states.
Because of parity symmetry, the upper surface spectrum with momenta
$(k_x,k_y)$ should be the same as that of the lower surface with
$(-k_x, -k_y)$.
Further, since the spin polarization is invariant under inversion operation, the corresponding surface states on the upper and lower surfaces are of opposite lattice helicity eigenvalues, thus only the surface states on the
lower surface are depicted in Fig. \ref{fig:spectra}.

The spectra are plotted in Fig. \ref{fig:spectra} along the high symmetry lines connecting time-reversal invariant points in the Brillouin zone.
In Fig. \ref{fig:spectra} ($a$), the spectra are presented along the cut of
$X_1$-$\Gamma$-$X_1$.
The Harper equation (Eq. (\ref{eq:harper_2})) along this cut is the same
as that of the 2D Hofstadter problem in the magnetic field
\cite{hatsugai1993} but with Kramers doubling because of the time-reversal
symmetry.
As a result, the surface modes appear in terms of Kramers doublet pairs.
For the band gap 1 to 6 counted from bottom to top, along the cut of
$X_1$-$\Gamma$-$X_1$, here are 3, 1, 2, 2, 1, and 3 branches
of Kramers doublets across the gap, respectively.
This pattern remains the same as along the cut from $M$-$\Gamma$-$M$
shown in Fig. \ref{fig:spectra} ($b$).
Along the path $M$-$X_1$-$M$, the surface states do not run across
the band gap.

\begin{table}[h]
\begin{tabular}{|c||c|c|c|c|c|c|c|}   \hline
$(k_x,k_y,k_z)$               & $B_1$ & $B_2$ & $B_3$ & $B_4$ & $B_5$
& $B_6$ & $B_7$
 \\ \hline  \hline
  $(0,0,0)$      & $+$ & $+$ & $-$ & $-$ & $+$ & $+$ & $-$ \\
  \hline
  $(0,0,\pi)$    & $+$ & $-$ & $+$ & $-$ & $+$ & $-$ & $+$ \\
  \hline
  $(0,\pi,0)$    & $+$ & $-$ & $+$ & $-$ & $+$ & $-$ & $+$ \\
  \hline
  $(0,\pi,\pi)$  & $-$ & $+$ & $+$ & $-$ & $-$ & $+$ & $+$ \\
  \hline
  $(\pi,0,0)$    & $+$ & $-$ & $+$ & $-$ & $+$ & $-$ & $+$ \\
  \hline
  $(\pi,0,\pi)$  & $-$ & $+$ & $+$ & $-$ & $-$ & $+$ & $+$ \\
  \hline
  $(\pi,\pi,0)$  & $+$ & $-$ & $+$ & $-$ & $+$ & $-$ & $+$ \\
  \hline
  $(\pi,\pi,\pi)$& $-$ & $+$ & $+$ & $-$ & $-$ & $+$ & $+$ \\
  \hline \hline
   $\prod_{a=1}^8 \xi_a^i$    & $-$ & $+$ & $-$ & $+$ & $-$ & $+$ & $-$ \\
\hline
   $\mathbb{Z}_2$  &$o$ & $o$ & $e$ & $e$& $o$ & $o$ &\\
\hline
\end{tabular}
\caption{The parity eigenvalues $\xi_a^i$ of the 3D bulk states at
eight time-reversal invariant states.
$B_i (i=1\sim 7)$ represents the $i$-th band counted from bottom to top,
and $a=1\sim 8$ is the index of states at the 8 time-reversal
invariant momenta.
The ${\mathbb Z}_2$ index $e$ or $o$ at the column of $B_i$ represents
that the system is ${\mathbb Z}_2$-trivial or non-trivial
when the lowest $i$ bands are filled.
The parameter values are $t_x=t_y=0.5$, $t_z=1$, and $p/q=2/7$.
}
\end{table}

\subsection{Boundary Fermi surface and the $\mathbb{Z}_2$-index}

To have a global view of the surface spectra, we present
the Fermi surfaces of states localized at the lower boundary of the system.
The Fermi energies lying between bands 2 and 3, and between
bands 3 and 4 are shown in Fig. \ref{fig:surfacecircle}
($a$) and ($b$), respectively.
In the former case, there is only one Fermi circle enclosing the
$\Gamma$ point,
which is helical according to the operator $\Sigma_{L}$, and thus
if bands 1 and 2 are filled, the system is $\mathbb{Z}_2$
topological non-trivial.
In this case,  the boundary Fermi surface can be viewed as a consequence
of a rotation of the edge Fermi points of the 2D QSH states in the $xz$ plane.
This is a strong 3D topological insulator, which means for
boundaries along the $xz$ and $yz$-directions, there should
also exist odd numbers of helical Fermi surfaces.

On the other hand, if the Fermi energy lies between bands 3 and 4,
there is one Fermi surface around each time-reversal invariant
point $\Gamma$, two $M$'s, and $R$, respectively.
Three of them are of the same helicity, and the other one is
of the opposite helicity.
For the case with bands 1, 2, and 3 filled, the
system becomes $\mathbb{Z}_2$ trivial.
We have also checked the cases in which the Fermi energy is lying in the gap
between bands 4 and 5, and between 5 and 6.
The topology of the helical boundary Fermi surfaces is very similar
to that depicted in Fig. \ref{fig:surfacecircle} ($a$) and ($b$),
respectively, but the helicity patterns are opposite.
Because bands 1 and 2 overlap, there does not exist a full direct gap over the entire Brillouin zone. The boundary Fermi surfaces are not plotted for the Fermi energy lying in this regime.

The lattice Hamiltonian of Eq. (\ref{eq:3dlattice}) conveniently enables the periodicity in the $z$-direction, so that we can quantitatively calculate its $\mathbb{Z}_2$-index.
Further, because of the inversion symmetry, the $\mathbb{Z}_2$-index is conveniently calculated by examining the parity eigenvalues of the bulk
states at time-reversal invariant lattice momenta \cite{fu2007}.
There are eight such lattice momenta $(k_x,k_y,k_z)$ with
$k_{x,y}=0$ or $\pi$, and $k_z=0$ or $\frac{\pi}{q}$.
The wave functions in each band at these momenta are parity
eigenstates and the associated eigenvalues $\xi^a_i$ are presented
in Table I, where $a=1\sim 8$ is the momentum index
and $i=1\sim 7$ is the band index.
For the case that the lowest $i$ bands are all filled, the $\mathbb{Z}_2$ index equals $\prod_{j=1}^i \prod_{a=1}^8 \xi^a_i$, which is presented
in Table I.
It shows that when the Fermi energy lies in the 1st, 2nd, 5th and 6th gaps, the system is $\mathbb{Z}_2$ non-trivial.
On the other hand, if the Fermi energy lies in the 3rd and the 4th gaps, the system is $\mathbb{Z}_2$ trivial, where the three weak $\mathbb{Z}_2$ indices are also trivial.
These results obtained from parity eigenvalues of bulk wave functions are consistent with that obtained from analyzing boundary helical Fermi surfaces in Fig. \ref{fig:surfacecircle}.

\begin{figure}[htbp]
\centering
\centering\includegraphics[width=0.49\linewidth]{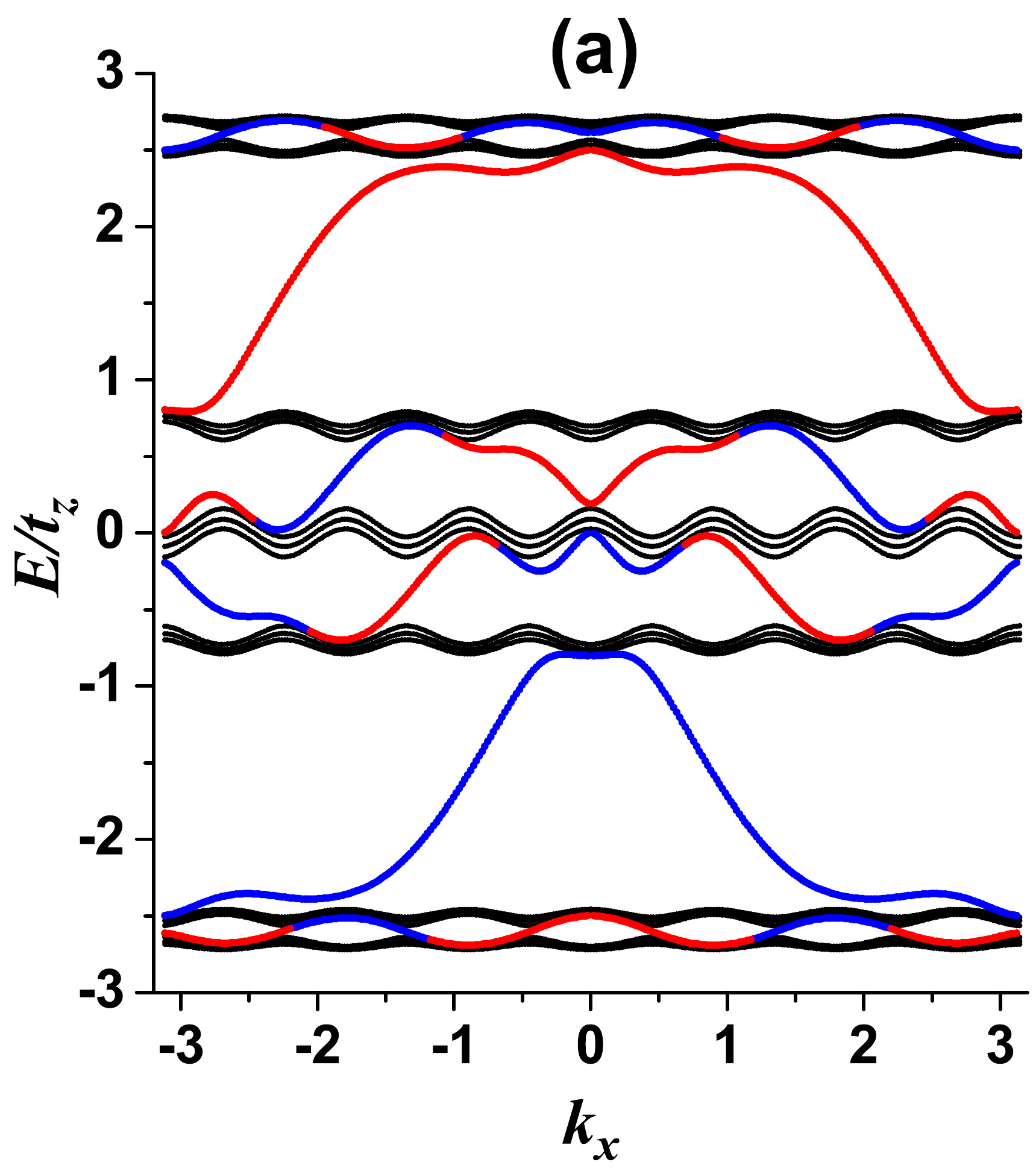}
\centering\includegraphics[width=0.49\linewidth]{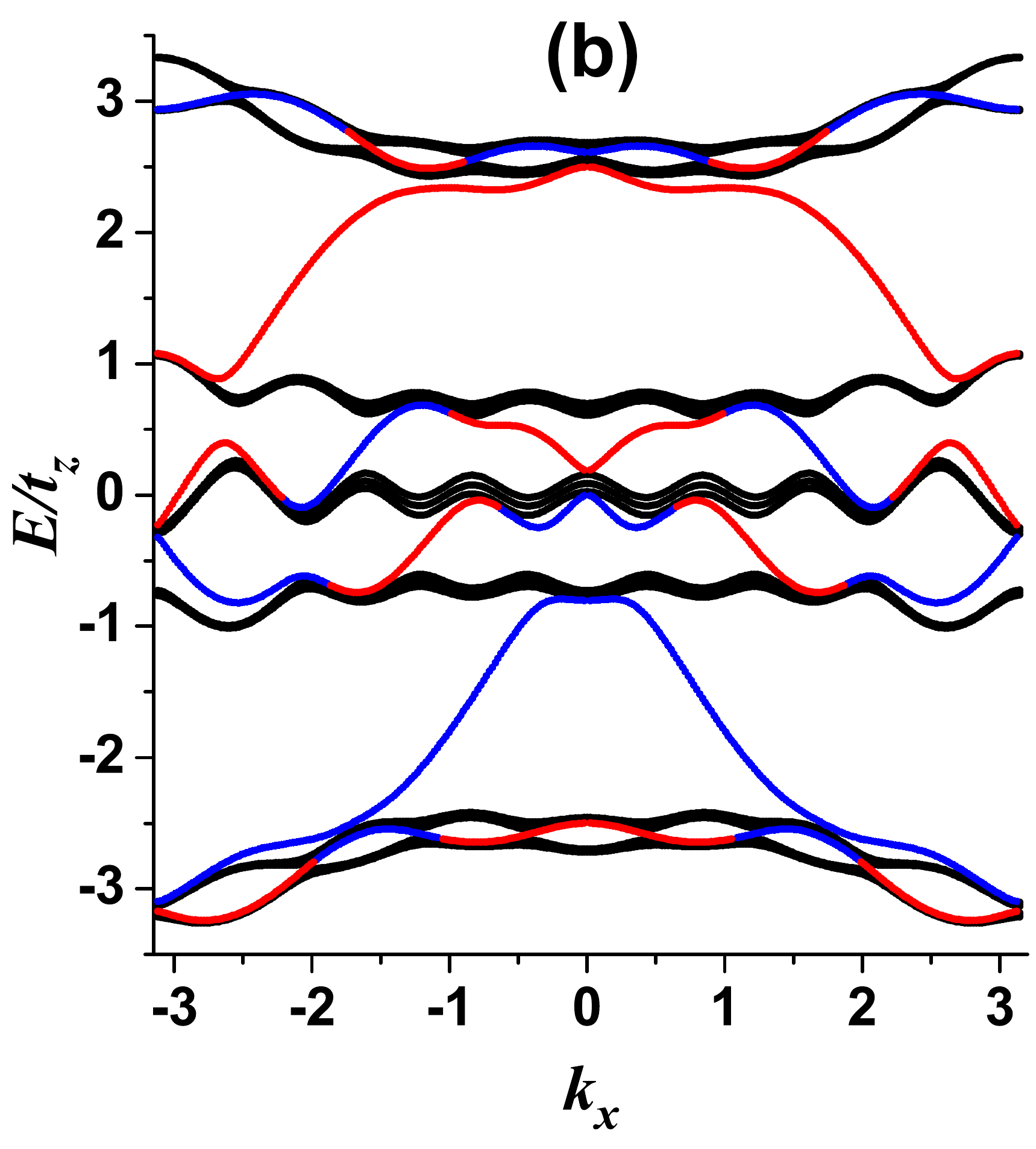}
\centering\includegraphics[width=0.49\linewidth]{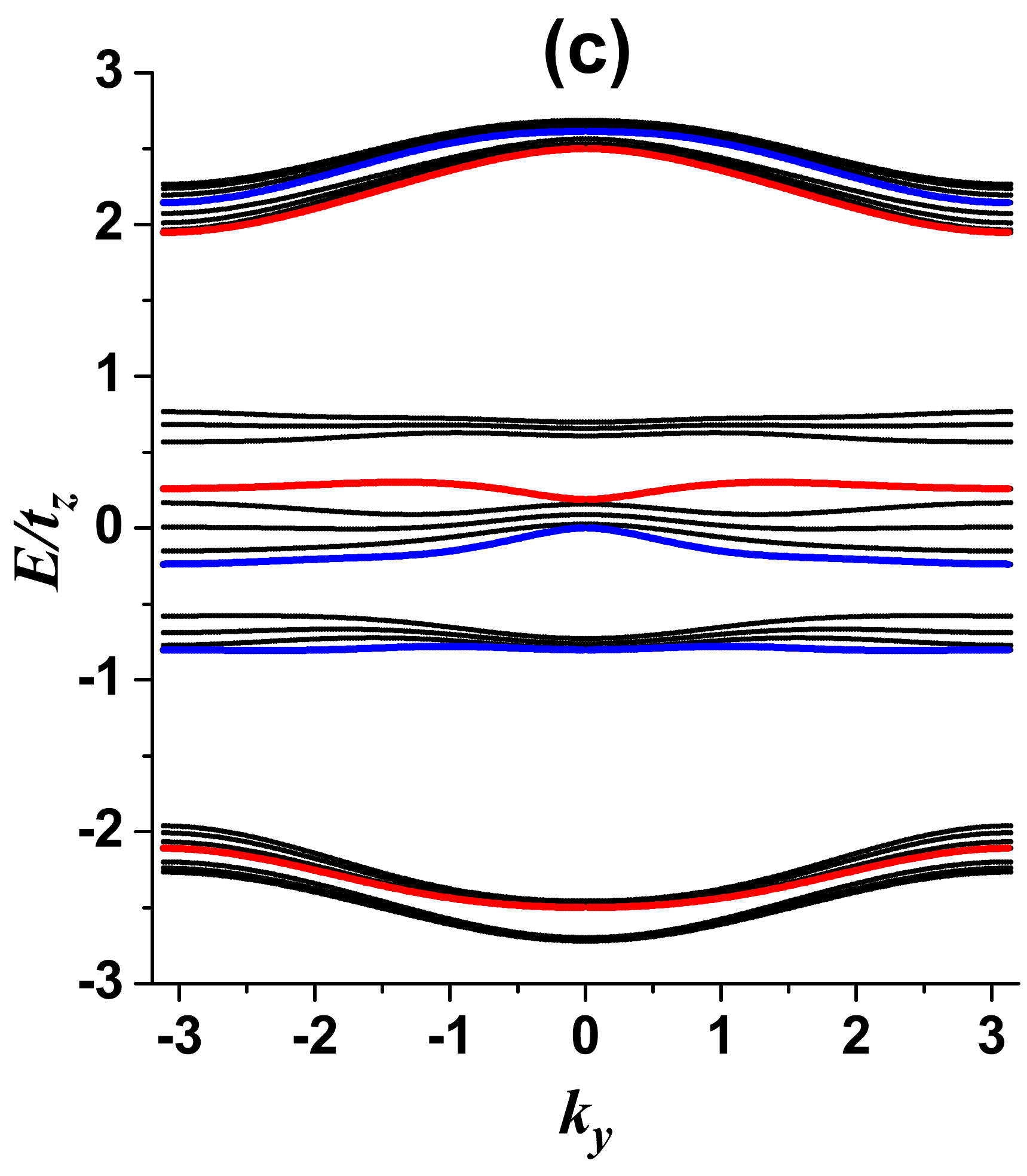}
\caption{The bulk and surface spectra along the time-reversal
invariant paths of ($a$) $X_1$-$\Gamma$-$X_1$, ($b$) $M$-$\Gamma$-$M$,
($c$)$X_2$-$\Gamma$-$X_2$, respectively.
The red and blue lines represent helical modes on the lower
boundary with positive and negative helicities, respectively.
The following parameters are used: $t_x=t_z=1$, $t_y=0.2$,
$p/q=2/7$, and $L_z=56$.
}
\label{fig:spectra_t02}
\end{figure}

\begin{figure}[htbp]
\centering
\centering\includegraphics[width=0.49\linewidth]{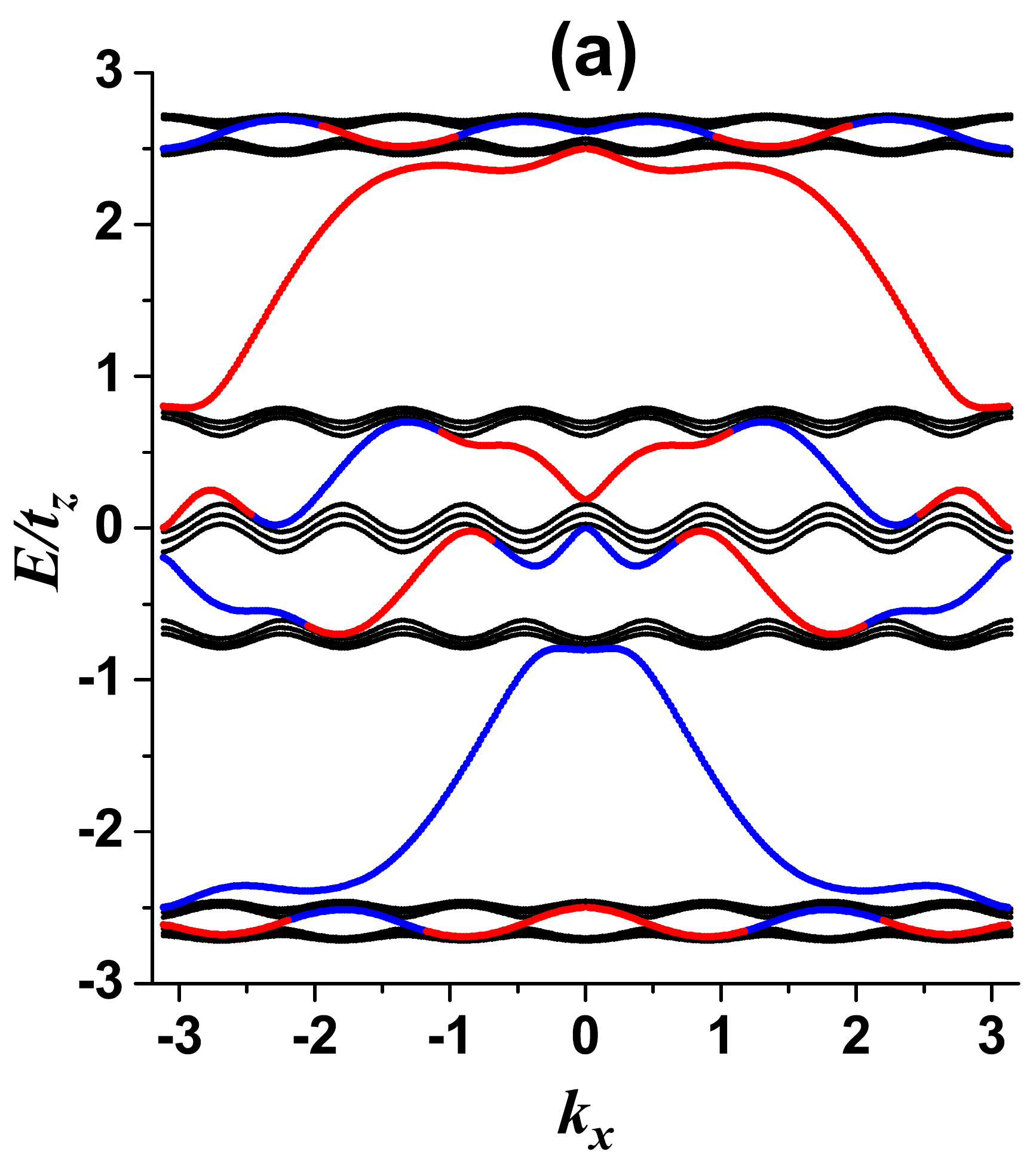}
\centering\includegraphics[width=0.49\linewidth]{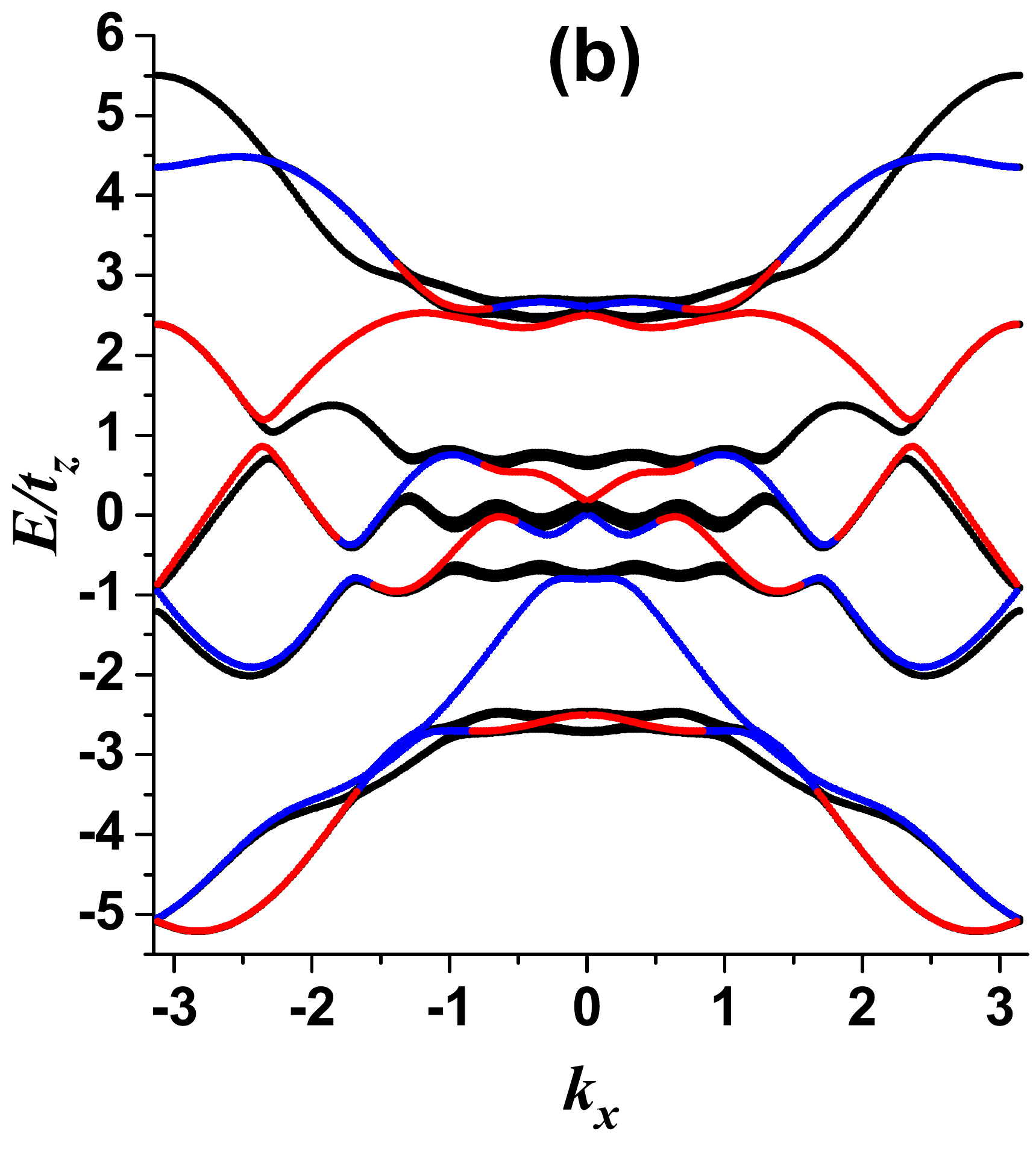}
\centering\includegraphics[width=0.49\linewidth]{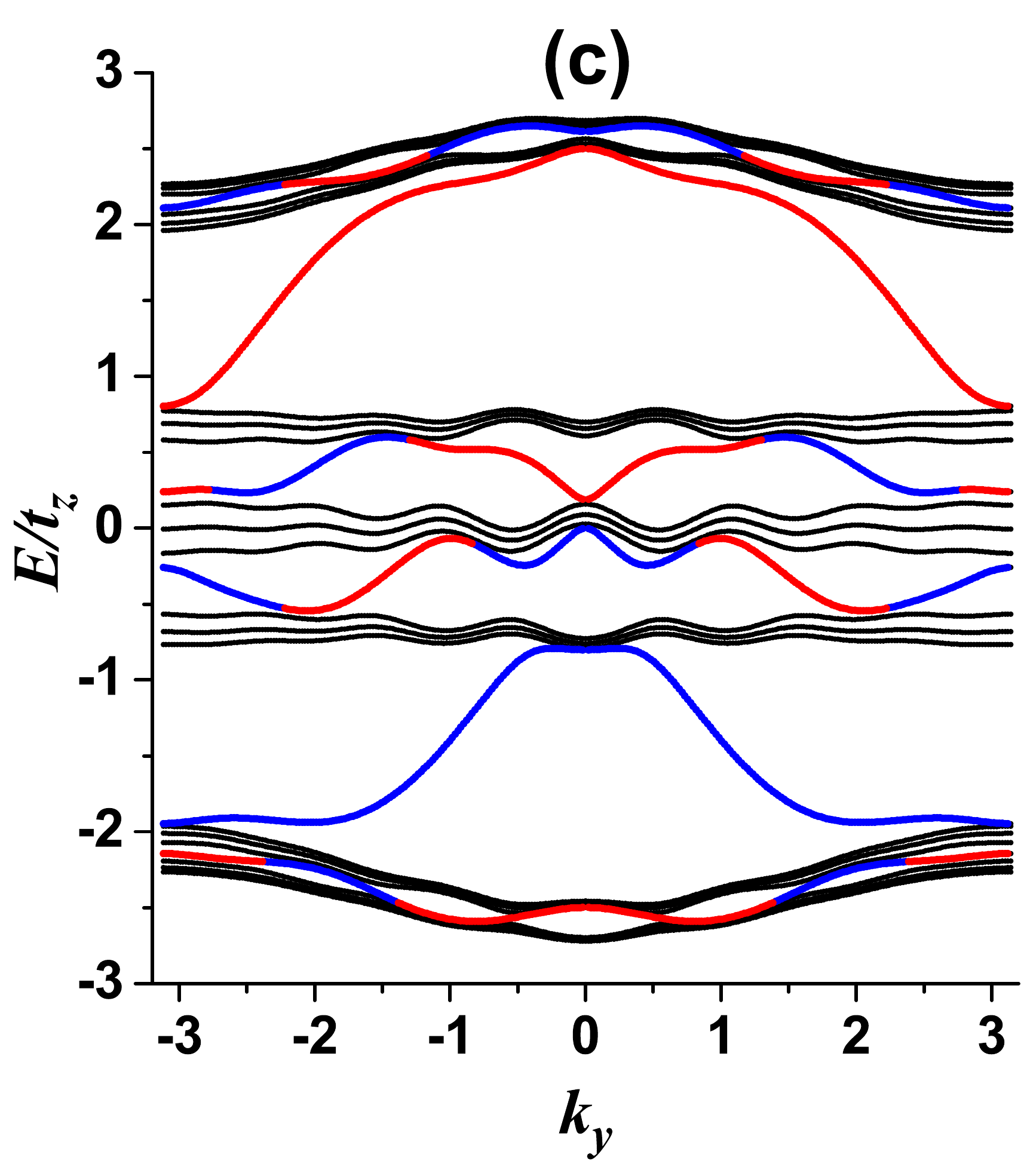}
\centering\includegraphics[width=0.49\linewidth]{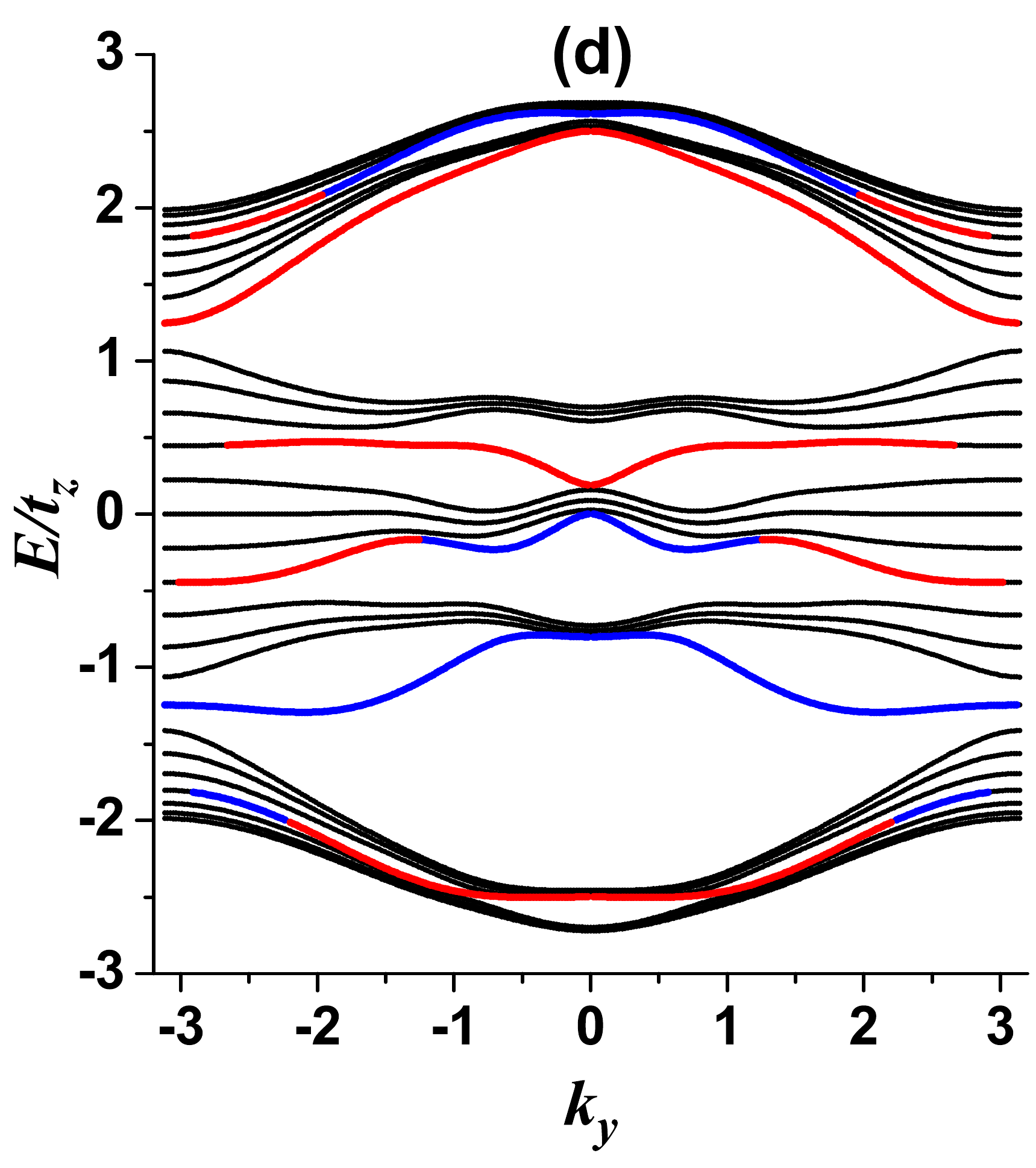}
\caption{The bulk and surface spectra along the time-reversal
invariant paths. ($a$) $X_1$-$\Gamma$-$X_1$, ($b$) $M$-$\Gamma$-$M$,
($c$)$X_2$-$\Gamma$-$X_2$ for $t_x=t_z=1$ and $t_y=0.8$;
($d$)$X_2$-$\Gamma$-$X_2$ for $t_x=t_z=1$ and $t_y=0.5$.
The red and blue lines represent helical modes on the lower
boundary with positive and negative helicities, respectively.
The following parameters are used:$p/q=2/7$, and $L_z=28$.
}
\label{fig:spectra_t08}
\end{figure}

\section{Transition from weak to strong topological insulating
states}
\label{sect:transition}

In this section, we consider the crossover of the Hofstadter problem
from the quasi-2D quantum spin Hall class to the 3D topological insulator
class, i.e, the transition between weak and strong $\mathbb{Z}_2$-classes.

\begin{table}
\begin{center}
\begin{tabular}{|c||c|c|c|c|c|c|c|}
  \hline
  $(k_x,k_y,k_z)$ & $B_1$ & $B_2$ & $B_3$ & $B_4$ & $B_5$ & $B_6$ & $B_7$  \\
  \hline \hline
  $(0,0,0)$      & $+$ & $+$ & $-$ & $-$ & $+$ & $+$ & $-$ \\
  \hline
  $(0,0,\pi)$    & $+$ & $-$ & $+$ & $-$ & $+$ & $-$ & $+$ \\
  \hline
  $(0,\pi,0)$    & $+$ & $+$ & $-$ & $-$ & $+$ & $+$ & $-$ \\
   \hline
  $(0,\pi,\pi)$  & $+$ & $-$ & $+$ & $-$ & $+$ & $-$ & $+$ \\
  \hline
  $(\pi,0,0)$    & $+$ & $-$ & $+$ & $-$ & $+$ & $-$ & $+$ \\
  \hline
  $(\pi,0,\pi)$  & $-$ & $+$ & $+$ & $-$ & $-$ & $+$ & $+$ \\
  \hline
  $(\pi,\pi,0)$  & $+$ & $-$ & $+$ & $-$ & $+$ & $-$ & $+$ \\
  \hline
  $(\pi,\pi,\pi)$& $-$ & $+$ & $+$ & $-$ & $-$ & $+$ & $+$ \\
  \hline\hline
  $\prod_{a=1}^8 \xi_a^i$    & $+$ & $+$ & $+$ & $+$ & $+$ & $+$ & $+$ \\
  \hline
  $\mathbb{Z}_2 $    & $e$ & $e$ & $e$ & $e$ & $e$ & $e$ &  \\
  \hline
  $\mathbb{Z}_{2,xz}$ &$o$ & $o$ & $e$ & $e$& $o$ & $o$& \\
  \hline
  $\mathbb{Z}_{2,yz}$ &$e$ & $e$ & $e$ & $e$& $e$ & $e$&  \\
  \hline
  $\mathbb{Z}_{2,xy}$ &$e$ & $e$ & $e$ & $e$& $e$ & $e$& \\
  \hline
\end{tabular}
\end{center}
\caption{The parity eigenvalues $\xi^i_a$ of the 3D bulk states at
eight time-reversal invariant states and values of the $\mathbb{Z}_2$
index.
Although the values of the strong $\mathbb{Z}_2$ index are all
trivial, those of the weak $\mathbb{Z}_2$ index can be non-trivial
in the $xz$-direction.
Parameter values are $t_x=t_z=1$, $t_y=0.2$, and $p/q=2/7$.
}
\end{table}

Now, let us set $t_x=t_z=1$ but tune $t_y$ from small values to 1.
In the case of $t_y=0$, Eq. \ref{eq:3dlattice} becomes decoupled
$xz$-layers. Each $xz$-layer is the Kramers doubling of the 2D Hofstadter problem with spin polarized along the $\pm y$-direction.
As $t_y$ increases, energy dispersion develops along the $y$-direction
and the direction of spin polarization also twists.
However, the band topology remains the same at small values of $t_y$.
In Fig. \ref{fig:spectra_t02} ($a$), ($b$), and ($c$), the bulk and
surface spectra at $t_y=0.2$ are presented along the cuts of $X_1$-$\Gamma$-$X_1$, $M$-$\Gamma$-$M$, and $X_2$-$\Gamma$-$X_2$, respectively.
Along $X_1$-$\Gamma$-$X_1$ and $M$-$\Gamma$-$M$, the bulk and surface spectra
are qualitatively the same as those in Fig. \ref{fig:spectra} ($a$)
and ($b$), respectively.
The numbers of branches of Kramers doubles are $3, 1, 2, 2, 1, 3$,
respectively, for the $i$-th band gap with $i=1\sim 6$.
However, the $xy$ plane is highly anisotropic.
Along the cut of $k_y$ axis, i.e., $X_2$-$\Gamma$-$X_2$, the spectra show
no non-trivial mid-gap boundary modes across all gaps.
In other words, this anisotropic case is a 3D weak topological
insulator, which is topologically non-trivial only along the
$xz$ plane.

This result can be confirmed from the calculation of the bulk
$\mathbb{Z}_2$-index.
Similarly to Table I, in Table II, we also present the corresponding parity eigenvalues
$\xi^a_i$ for bulk wave functions at eight time-reversal invariant momenta
($a=1\sim 8$) of bands $i$ ($i=1\sim 7$).
The values of the strong $\mathbb{Z}_2$-index are all trivial for
all the band gaps.
Nevertheless, the weak $\mathbb{Z}_2$-index in the $xz$-direction
exhibits a non-trivial pattern, which is non-trivial for
the $i$-th band gap with $i=1,2,5$ and $6$, and thus is
consistent with the evenness of the branches of boundary
Kramers doublets presented in Fig. \ref{fig:spectra_t02} ($a$).
The weak $\mathbb{Z}_2$ indices for the $xy$ and
$yz$-directions are all trivial.
Therefore, at $t_y=0.2$, the system is qualitatively stacked by a 2D-like
topological insulator in the $xz$ plane, and it forms a 3D weak topological insulator.

Now let us study the case with a large value of $t_y=0.8$.
The bulk and boundary spectra are plotted in Fig.
\ref{fig:spectra_t08} ($a$), ($b$) and ($c$)
along the same cuts as taken in Fig. \ref{fig:spectra_t02}.
In particular, along the $y$-direction cut $X_2$-$\Gamma$-$X_2$ in
Fig. \ref{fig:spectra_t08} ($c$), non-trivial boundary modes
appear across the band gap, which exhibit a pattern similar to the spectra along the $x$-direction cut $X_1$-$\Gamma$-$X_1$.
Although small anisotropy remains in the $xy$ plane, the system
has already become a 3D strong topological insulator.
We have also calculated the parity eigenvalues of bulk
wave functions at time-reversal invariant momenta,
and the results are the same as that in Table I, which also
indicates that the strong $\mathbb{Z}_2$ index is non-trivial.

The transition between
weak and strong topological insulating states as $t_y$ varies must involve bulk band closing. It can be found that the transition point occurs at $t_y=0.5$.
In this case, the periodic spin-dependent potential characterizing the
Hofstadter problem $V_{n_z;ss^\prime}$ defined in
Eq. (\ref{eq:spin_incom}) vanishes at $\vec k_{2D}=(0,\pi)$.
As shown in Fig. \ref{fig:spectra_t08} ($d$), all the
band gaps closes at $(0,\pi)$, which triggers this
band topology transition.

\section{Discussions}
\label{sect:disc}

The above study of SU($2$) Hofstadter problems in three-dimensions differs fundamentally
from previously studied 3D Hofstadter problems under a tilted magnetic field
\cite{halperin1987,montambaux1990,kohmoto1992,koshino2001,koshino2004}.
Firstly, the latter problem is U($1$) in nature. There is only one fermion component on each
lattice site; the phase on each link and the flux penetrating each plaquette are
U(1). However, our case is a 3D non-Abelian SU($2$) problem. Here, the spin-1/2 spinor is located on each lattice site, and the gauge potentials defined on links are noncommutative SU(2) phases.
Secondly, the bulk topological classes and surface spectra of these
two problems are fundamentally different.
Our SU(2) case maintains time-reversal symmetry, and hence it belongs to the
AII class characterized by the $\mathbb{Z}_2$-indices calculated above
\cite{ryu2010}.
When the strong $\mathbb{Z}_2$-index is odd, it is a strong 3D
$\mathbb{Z}_2$ topological insulator, and the surface spectra
exhibit an odd number of Dirac cones as shown in Fig. \ref{fig:surfacecircle}.
In contrast, the 3D U(1) problem breaks time-reversal symmetry, and
it is not a genuinely novel topological class.
Roughly speaking, it belongs to the AI class and can be understood the same as
 stacking layers of 2D quantum Hall systems.
In other words, it is a weak 3D topological state inheriting 2D features.
The boundary Fermi surfaces do not close form Dirac cones but rather
disconnect forming Fermi arcs.
Finally, the reduced 1D Hofstadter equations are also different in these two cases.
Our Eq. \ref{eq:harper} is in fact a matrix equation in spin space, and
the associated transfer matrix Eq. \ref{eq:transfer_1} is
off-diagonal on spin-indices.
The partially diagonalized transfer matrix Eq. \ref{eq:transfer_2}
is defined in the eigenbasis of the helicity operator $\Sigma_L$.
This set of helical eigenbasis render the non-trivial
$\mathbb{Z}_2$ topological properties.

Recently, a paper\cite{ganeshan2015} considered the 3D Hofstadter
problem which corresponds to a tilted magnetic field in the cubic
lattice with both the nearest and next nearest neighbor hopping.
When the flux of each plaquette equals $\pi$, time-reversal
symmetry is restored.
The reduced 1D Aubry-Andre-Harper model belongs to the BDI class
\cite{ryu2010} and the bulk spectra show Weyl points.
Again the flux in this model is complex, and the realized time-reversal symmetry
satisfies $T^2=1$; while, in our case the time-reversal symmetry satisfies $T^2=-1$ because of the Kramers degeneracy based
on the SU(2) structure.
In other words, the time-reversal symmetry in our case is in
in the symplectic class; while, it is in the orthogonal class
in Ref. \onlinecite{ganeshan2015}.

Next we discuss possible experimental realizations of the 3D
SU(2) Hofstadter problem.
In Refs. \onlinecite{li2013, li2013a}, the 3D SU(2) Landau
level problems in the continuum have been proposed to be
realized in strained semiconductor systems.
Nevertheless, as in the case of the U(1) Hofstadter
problem, the characteristic length scale of Landau level
states is too long compared with the lattice constant
and thus the lattice effect can be neglected.
The recent development of ultra-cold atom physics provides a
promising opportunity to realize the SU(2) Hofstadter problem.
Based on the available experimental techniques, such as
laser assistant tunneling and shaken lattices, which
realized the U(1) Hofstadter problem \cite{miyake2013,
aidelsburger2013,celi2014,jaksch2003}, it is natural to expect that the
3D SU(2) Hofstadter problem can also be studied
experimentally in the near future.
Basically, these techniques need to be extended to spin-dependent
hopping or shaking.
For a given direction, the associated spin eigenstates
are addressed oppositely, such that they acquire opposite
phases maintaining time-reversal symmetry.
In principle, different spin eigenstates correspond to
different internal states of atoms, which have different
responses to external lasers, and thus the above spin-dependent
manipulations can be realized.
The key difficulty is that spin eigen-directions are orthogonal
along $x$ and $y$-bonds due to their non-Abelian nature.
These could be realized by different sets of laser beams that
control hoppings along the $x$ and $y$ directions separately,
and their effects could be superposed together.
Along each direction, the spin eigenstates are chosen
according to orthogonal Pauli matrices, respectively.
Nevertheless careful designs are still needed to achieve
the desired SU(2) Hofstadter Hamiltonian.

\section{Conclusions}
\label{sect:conclusions}

In summary, we have constructed the time-reversal invariant 3D Hofstadter problem based on the 3D SU(2) Landau levels in the Landau-type gauge.
This lattice model provides an $SU(2)$ generalization of the usual 2D Hofstadter problem in a magnetic field. For each pair of in plane momenta $(k_x, k_y)$, this system is reduced to 1D described by a generalized $SU(2)$ Harper equation with a helicity structure.

Different from its continuum version, this lattice system possesses 3D translation symmetry as characterized by the periodic spin-dependent potential in the $SU(2)$ Harper equation. Hence, quantitative analysis of the nontrivial $\mathbb{Z}_2$ band topology is performed.
In the energy spectra studied in this paper, boundary states with opposite lattice helicity structures are spatially separated at different boundaries; $\mathbb{Z}_2$ non-trivial helical boundary Fermi surface are found and consistent with the $\mathbb{Z}_2$ bulk band topology analyzed based on the parity eigenvalues of the bulk wave functions.
The transition of the band topology from a weak topological
insulator to a strong one is also investigated.

\acknowledgments
Y. L. thanks D. Arovas, T. L. Ho, W. Ketterle, and C. Wu for helpful discussions.
Y. L. acknowledges  the support at the Princeton Center for Theoretical Science.
Y. L. also thanks T. L. Ho for his hospitality.

%

\end{document}